\documentclass[9pt,twocolumn,twoside]{osajnl}

\journal{ol} 

\setboolean{shortarticle}{true}
\newcommand{\orcid}[1]{\href{https://orcid.org/#1}{\resizebox{10px}{!}{\includegraphics{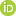}}}}
\title{Predictability, Distinguishability and Entanglement}

\author[]{Tabish Qureshi \orcid{0000-0002-8452-1078}}

\affil[]{Centre for Theoretical Physics, Jamia Millia Islamia, New Delhi-110025, India.}

\affil[]{Email: tabish@ctp-jamia.res.in}

 \doi{\url{http://dx.doi.org/10.1364/OL.415556}}

\begin{abstract}
Recent times have seen a spurt of research activity focused on ``completing"
certain wave-particle duality relations using entanglement or polarization.
These studies use a duality relation involving path-predictability, and not
path-distinguishability. Quantum origins of these results are explored here,
in the more general framework of multipath quantum interference. Multipath
interference with a path-detector is theoretically analyzed to find
the connection between predictability and distinguishability. It is shown
that entanglement is what quantitatively connects distinguishability with
predictability. Thus, a \emph{duality} relation between distinguishability and
coherence, can also be viewed as a \emph{triality} between predictability,
entanglement and coherence. There exist two different kinds of duality
relations in the literature, which pertain to two different kinds of
interference experiments, with or without a path-detector. Results of
this study show that the two duality relations are quantitatively connected 
via entanglement. The roots of the new results in the classical optical
domain, including the polarization coherence theorem, can be understood
in the light of this work. 
\end{abstract}

\setboolean{displaycopyright}{true}

\begin{document}

\maketitle

Wave-particle duality is an old subject in quantum physics \cite{bohr},
and has seen a sustained interest, which has only increased with time.
Two-path interference became a testing ground for all such ideas. Greenberger
and Yasin looked at the issue of wave-particle duality from a simplified
perspective where the ``particleness" is inferred only if the quanton
is more likely to follow one path than the other. For such a scenario,
they derived the inequality \cite{greenberger,jaeger}
\begin{equation}
\mathcal{P}^2 + \mathcal{V}^2 \le 1,
\label{GY}
\end{equation}
where $\mathcal{P}$ is called path-predictability, or just predictability,
and $\mathcal{V}$ is the interference visibility, a measure of the wave
nature of the quanton. In a two-path experiment, the states corresponding
to the quanton passing through different paths, are necessarily orthogonal,
and they can be used as basis states of an effectively 2-dimensional
Hilbert space. The density matrix of the quanton can be written in this
basis, and the diagonal elements $\rho_{11}, \rho_{22}$ represent the
probabilities of the quanton to
pass through path 1 and 2, respectively. Predictability is defined as
\cite{greenberger,jaeger}
\begin{equation}
\mathcal{P} = |\rho_{11} - \rho_{22}| = \sqrt{1-4\rho_{11}\rho_{22}}.
\label{P2}
\end{equation}
It is obvious that if the two beams are of same intensity, $\mathcal{P} = 0$.
The visibility is defined simply as the standard fringe contrast
\begin{equation}
\mathcal{V} = \frac{I_{max}-I_{min}}{I_{max}+I_{min}},
\label{V}
\end{equation}
where $I_{max}, I_{min}$ are the maximum and minimum intensities,
respectively.
The inequality Eq. (\ref{GY}) becomes an equality if the quanton state
is pure, but if the quanton state is not pure due a variety of reasons,
the inequality holds. On the other hand,
if one wants to experimentally know which of the two paths the
quanton followed, even if it is equally likely to go through any of the
two, one has to have a path-detector in place, to make the two paths
distinguishable \cite{wootters}.
Englert derived the following duality relation for such a
situation \cite{englert}
\begin{equation}
\mathcal{D}^2 + \mathcal{V}^2 \le 1,
\label{englert}
\end{equation}
where $\mathcal{D}$ is called path-distinguishability, or just
distinguishability. Here too, the inequality becomes an equality if the
quanton and path-detector state is pure. 
Wave-particle duality has also been widely explored in classical optics.
It should be emphasized here that in the classical optical scenario,
the duality relation of the kind Eq. (\ref{GY}) is meaningful. Since a 
question like, which of the two paths did the quanton actually follow,
are not meaningful in the classical optical realm, formulation of
a duality relation of the kind Eq. (\ref{englert}) is not needed there.

Recently it was shown that, in the realm of classical optics, including
polarization in the two-slit interference experiment, the following
``triality" relation can be derived \cite{eberly3}
\begin{equation}
D^2 + V^2 + C^2 = 1,
\label{trialityC}
\end{equation}
where $D$ is what the authors call ``distinguishability," but should actually
be identified with the predictability $\mathcal{P}$, $V$ is the interference
visibility given by Eq. (\ref{V}) and $C$ is \emph{concurrence}, a measure of
entanglement. The
authors interpret this equality as ``completing the duality relation"
given by Eq. (\ref{GY}), which is an inequality. This work was built upon their
earlier work recognizing entanglement like non-separability involving
polarization or a generalized polarization, which has important consequences
in classical optics \cite{eberly1,eberly2}, and also upon the recognition
of the role of entanglement in complementarity relations in a bipartite
system of two qubits \cite{jakob}. The relationship between polarization,
indistinguishability, and entanglement has also been probed earlier
\cite{zela2}. This work was followed by a flurry of papers on similar theme
\cite{copt1,copt2,copt3,eberly4,pctexpt,zela1}.
These results point to the interesting role of entanglement in wave-particle
duality. Extending these ideas to the level of single photons should throw
more light on the quantum origin of these results. That is the aim of
the present investigation. In particular, we consider multipath quantum
interference in the presence of a path detector.

\begin{figure}
\centering
\includegraphics[width=8.0 cm]{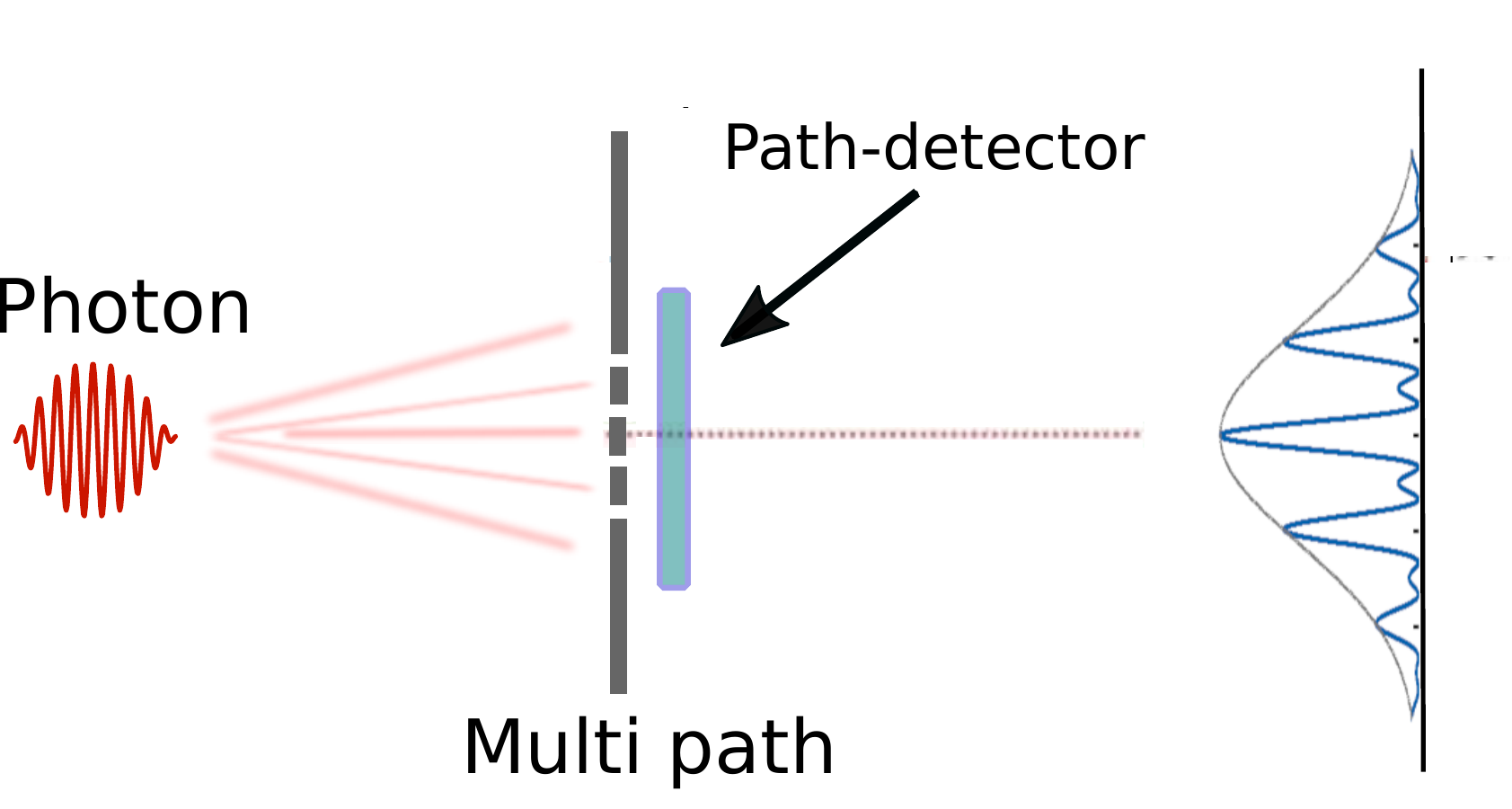}
\caption{Schematic representation of a n-path interference experiment
with a path-detector.}
\label{npath}
\end{figure}

Let us consider a quanton passing through a multipath arrangement, and
there is a path-detector present whose precise nature we need
not specify for the present purpose (see FIG. \ref{npath}).
We start by specifying a general pure state of a quanton passing through
a $n$-path interferometer. If $|\psi_k\rangle$ represent the
state corresponding to the quanton taking the k'th path, 
the initial state, before the quanton interacts with the path-detector,
is given by
\begin{equation}
|\Psi_0\rangle = c_1|\psi_1\rangle + c_2|\psi_2\rangle + 
\dots + c_n|\psi_n\rangle ,
\label{Psi0}
\end{equation}
where $|c_k|^2$ represents the probability of the quanton taking the k'th path.
The states $\{|\psi_i\rangle\}$ can be assumed to form an ortho-normal set,
without loss of generality.
The combined state of the quanton and the path-detector, as the quanton
emerges from the multipath arrangement, can be represented as
\begin{equation}
|\Psi\rangle = c_1|\psi_1\rangle|d_1\rangle + c_2|\psi_2\rangle|d_2\rangle + 
\dots + c_n|\psi_n\rangle|d_n\rangle ,
\label{Psi}
\end{equation}
where $\{|d_i\rangle\}$ represent certain normalized states of the
path-detector which may not necessarily be orthogonal to each other.
This entanglement is a fundamental requirement of the process of measurement,
as laid down by von Neumann \cite{neumann}.

In the absence of the path-detector, one can derive a duality relation
\cite{roy}
\begin{eqnarray}
\mathcal{P}^2 + \mathcal{C}^2 = 1 .
\label{pc}
\end{eqnarray}
where $\mathcal{P}$ is a \emph{generalized predictability} 
\begin{equation}
\mathcal{P} \equiv \sqrt{1 - \Big(\tfrac{1}{n-1}\sum_{j\neq k}
\sqrt{\rho_{jj}}\sqrt{\rho_{kk}} \Big)^2},
\label{P}
\end{equation}
$\rho$ being the density operator for the pure quanton state (\ref{Psi0}),
$\rho=|\Psi_0\rangle\langle\Psi_0|$, and the basis states are taken to be
$\{|\psi_i\rangle\}$. For $n=2$, Eq. (\ref{P}) reduces to Eq. (\ref{P2}).
The wave-nature is quantified by the recently introduced
measure of coherence \cite{coherence,tqcoherence}
\begin{equation}
\mathcal{C} \equiv \frac{1}{n-1}\sum_{j\neq k} |\rho_{jk}| ,
\label{C}
\end{equation}
where $\rho$ is the density operator of the quanton, and the basis is
chosen to be $\{|\psi_i\rangle\}$. This duality relation (\ref{pc}) is
the generalization of Greenberger and Yasin's relation (\ref{GY}) to
$n$ paths. For $n=2$, the $\mathcal{P}$ and $\mathcal{C}$ of Eq. (\ref{pc})
reduced to the $\mathcal{P}$ and $\mathcal{V}$ of Eq. (\ref{GY}), respectively.
If the quanton experiences some incoherence because of some reasons,
Eq. (\ref{pc}) becomes an inequality.

Interestingly, a simpler form of predictability can be formulated, as has
been shown earlier in the classical case \cite{opticoh}. This simpler
predictability has the following form
\begin{equation}
\mathcal{P}_Q \equiv 1 -\tfrac{1}{n-1}\sum_{j\neq k}
\sqrt{\rho_{jj}}\sqrt{\rho_{kk}}.
\label{PQ}
\end{equation}
This yields a simple and interesting duality relation
\begin{eqnarray}
\mathcal{P}_Q + \mathcal{C} = 1 .
\label{pqc}
\end{eqnarray}

Now we move on to the more interesting case where there \emph{is} a
path-detector in place. For the state (\ref{Psi}), the reduced density
operator of the quanton is given by
$\rho_r=\sum_{i=1}^n\sum_{j=1}^n c_ic_j^* \langle d_j|d_i\rangle \rho_{ij}$,
where $\rho$ is the density matrix of the pure state (\ref{Psi0}), same as
that appears in Eq. (\ref{P}) and Eq. (\ref{PQ}).
For this case one can define a \emph{path-distinguishability} as
\cite{3slit,cd15}
\begin{eqnarray}
\mathcal{D}_Q &\equiv& 1 - \tfrac{1}{n-1}\sum_{i\neq j}
\sqrt{\rho_{ii}\rho_{jj}} |\langle d_i|d_j\rangle|
\label{DQ}
\end{eqnarray}
This distinguishability is based on unambiguous quantum state discrimination
(UQSD) \cite{uqsd1,uqsd2,uqsd3,uqsd4}. It is essentially the maximum
probability with which the $n$ paths can be distinguished \emph{unambiguously}.
With the reduced density matrix, the coherence is modified to
\begin{equation}
\mathcal{C} = \tfrac{1}{n-1}\sum_{j\neq k} |\rho_{jk}|\ |\langle d_i|d_j\rangle| .
\label{Cn}
\end{equation}
We wish to emphasize here that the coherence $\mathcal{C}$, given by 
Eq. (\ref{Cn}) is a good measure of the wave-nature. Just like conventional
visibility, it can be measured in an interference experiment 
\cite{tqcoherence,tania}.
Using these definitions, a tight multipath wave-particle duality relation has
been derived before \cite{cd15}
\begin{eqnarray}
\mathcal{D}_Q + \mathcal{C} = 1 .
\label{dqc}
\end{eqnarray}
This relation is an equality even though the quanton is entangled with the
path-detector. It can also be cast in a quadratic form similar to that of
Eq. (\ref{englert}), by defining a different distinguishability, as
$\mathcal{D}=\sqrt{\mathcal{D}_Q(2-\mathcal{D}_Q)}$. With this new
distinguishability, one can write the tight duality relation \cite{nduality}
\begin{equation}
\mathcal{D}^2 + \mathcal{C}^2 = 1,
\label{nduality}
\end{equation}
which should be viewed as the multipath generalization of Eq. (\ref{englert}).
It should be mentioned here that for $n=2$, this $\mathcal{D}$ reduces to
the $\mathcal{D}$ of Eq. (\ref{englert}), and Eq. (\ref{nduality}) reduces to
Eq. (\ref{englert}).

Relations (\ref{pc}) and (\ref{pqc}) involve predictability, and pertain to
an experiment without any path-detector. Relations (\ref{nduality}) and
(\ref{dqc}) involve distinguishability, and pertain to interference 
experiments where there is a path-detector. Could there be a connection
between the two? From Eq. (\ref{PQ}) and Eq. (\ref{DQ}), path-distinguishability
may be written as
\begin{eqnarray}
\mathcal{D}_Q &=& 1 - \tfrac{1}{n-1}\sum_{i\neq j}\sqrt{\rho_{ii}\rho_{jj}}\nonumber\\
&+& \tfrac{1}{n-1}\sum_{i\neq j}
\big(\sqrt{\rho_{ii}\rho_{jj}} - \sqrt{\rho_{ii}\rho_{jj}}
|\langle d_i|d_j\rangle|\big) \nonumber\\
&=& \mathcal{P}_Q + \mathcal{E}_Q .
\label{DPQ}
\end{eqnarray}
Now, the quantity $\mathcal{E}_Q$ is interesting, and can be written as follows
\begin{eqnarray}
\mathcal{E}_Q &=& \tfrac{1}{n-1}\sum_{i\neq j}
\big(\sqrt{\rho_{ii}\rho_{jj}} - \sqrt{\rho_{ii}\rho_{jj}}
|\langle d_i|d_j\rangle|\big) \nonumber\\
&=& \tfrac{1}{n-1}\sum_{i\neq j}
\big(\sqrt{{\rho_r}_{ii}{\rho_r}_{jj}} - |{\rho_r}_{ij}| \big),
\label{EQ}
\end{eqnarray}
where $\rho_r$ is the reduced density operator for the state $|\Psi\rangle$,
given by Eq. (\ref{Psi}). Notice that if all $|d_i\rangle$ are identical, i.e.,
the state $|\Psi\rangle$ is disentangled, $\mathcal{E}_Q=0$. If all
$|d_i\rangle$ are mutually orthogonal, and $|c_i|=1/\sqrt{n}$, for all
$i$, which means the state is maximally entangled, $\mathcal{E}_Q=1$.
Thus $\mathcal{E}_Q$ can be considered a measure of entanglement. It is
interesting to compare $\mathcal{E}_Q$ to a well known entanglement measure,
\emph{I-concurrence} defined as \cite{rungta,pani}
\begin{eqnarray}
E_c^2 &=& 2\sum_{i\neq j}
\left({\rho_r}_{ii}{\rho_r}_{jj} - |{\rho_r}_{ij}|^2 \right).
\label{Icon}
\end{eqnarray}
One can see that $\mathcal{E}_Q$ is a measure of purity of $\rho_r$ in
the same way as $E_c$. The RHS of Eq. (\ref{Icon}) involves elements of a 
$2\times 2$ sub-matrix specified by $i$ and $j$, summed over all $i,j$. As all
terms in the sum are positive-definite, a decrease
in the entanglement measure $E_c^2$ can only be caused by the term
$|{\rho_r}_{ij}|^2$ becoming closer to ${\rho_r}_{ii}{\rho_r}_{jj}$,
in one or more submatrices. But this also means that the term
$|{\rho_r}_{ij}|$ becomes closer to $\sqrt{{\rho_r}_{ii}{\rho_r}_{jj}}$,
which means that the measure $\mathcal{E}_Q$ defined by Eq. (\ref{EQ}) 
decreases. By this argument, whenever $E_c^2$ decreases, $\mathcal{E}_Q$
will also decrease. Thus $\mathcal{E}_Q$ is expected to show the same
monotonicity as $E_c$. This indicates that $\mathcal{E}_Q$ possesses the 
properties of a measure of entanglement, that may be sufficient for
the present purpose. However, a more rigorous analysis may be needed to
qualify it as a proper entanglement measure. So, we can say that the relation
\begin{equation}
\mathcal{D}_Q = \mathcal{P}_Q + \mathcal{E}_Q
\end{equation}
shows that distinguishability is connected to predictability in a
precise way through entanglement. The duality relation (\ref{dqc})
can now be written as
\begin{eqnarray}
\mathcal{P}_Q + \mathcal{C} + \mathcal{E}_Q = 1 ,
\label{pceq}
\end{eqnarray}
which is a triality relation between predictability, quantum coherence
and entanglement.

A similar analysis can be carried out for the quadratic form of the duality
relation (\ref{nduality}). Here we write $\mathcal{D}$ in terms of
$\mathcal{P}$ using Eq. (\ref{P}):
\begin{eqnarray}
\mathcal{D}^2 &=& \mathcal{D}_Q(2 - \mathcal{D}_Q) =
1 - \big(\tfrac{1}{n-1}\sum_{i\neq j}\sqrt{\rho_{ii}\rho_{jj}}
|\langle d_i|d_j\rangle|\big)^2\nonumber\\
&=& 
1 - \big(\tfrac{1}{n-1}\sum_{i\neq j}\sqrt{\rho_{ii}\rho_{jj}}\big)^2\nonumber\\
&+& \tfrac{1}{(n-1)^2}\Big[ \big( \sum_{i\neq j}
 \sqrt{\rho_{ii}\rho_{jj}}\big)^2 - \big(\sum_{i\neq j}
\sqrt{\rho_{ii}\rho_{jj}} |\langle d_i|d_j\rangle|\big)^2\Big] \nonumber\\
&=& \mathcal{P}^2 + \mathcal{E}^2 ,
\label{DPC}
\end{eqnarray}
where
\begin{eqnarray}
\mathcal{E}^2 = \tfrac{1}{(n-1)^2}\Big[ \big( \sum_{i\neq j}
 \sqrt{\rho_{ii}\rho_{jj}}\big)^2 - \big(\sum_{i\neq j}
\sqrt{\rho_{ii}\rho_{jj}} |\langle d_i|d_j\rangle|\big)^2\Big] 
\label{Esq}
\end{eqnarray}
can also be considered an indicator of entanglement using an argument similar
to that used for $\mathcal{E}_Q$. The confidence in this argument is
reinforced by the fact that for $n=2$, $\mathcal{E}$
reduces to the I-concurrence $E_c$ given by Eq. (\ref{Icon}). For $n>2$,
$\mathcal{E}$ is bounded between 0 and 1, whereas $E_c$ is not. Thus
we arrive at another triality relation
\begin{eqnarray}
\mathcal{P}^2 + \mathcal{C}^2 + \mathcal{E}^2 = 1 ,
\label{pce}
\end{eqnarray}
between predictability, coherence and entanglement. This relation may
be looked upon as a multipath, quantum version of Eq. (\ref{trialityC}).
It should rather be looked upon as the root from which relations
like Eq. (\ref{trialityC}) arise, because the derivation is fully quantum,
in the sense that we started from a pure quantum state which could apply
to single quantum particles.
The entanglement in it is also genuinely quantum. In the classical
optical regime, it is not genuine entanglement, but non-separability
of two degrees of freedom of the light field \cite{karimi}. Similar ideas
have recently been explored in a somewhat different way \cite{maziero}.

This result is profound as it tells us that entanglement is what connects
distinguishability to predictability, something that was not recognized
earlier. It tells us that the pairs of relations (\ref{GY})
and (\ref{englert}), (\ref{pc}) and (\ref{nduality}), and 
(\ref{pqc}) and (\ref{dqc}), are not unconnected duality relations. One
is connected to the other through entanglement. For example, if in a
multipath interference, the entanglement measure (between the quanton
and path detector) goes to zero, one smoothly goes from (say) Eq. (\ref{nduality})
to Eq. (\ref{pc}).

Connecting distinguishability to predictability is just one aspect of
this analysis. The role of entanglement is more than that. If the
system, the quanton is entangled with, is part of the path-detecting
device, then of course entanglement is useful in relating distinguishability
to predictability. But what if the quanton is entangled to a system
which is not part of the path-detecting apparatus? In such a situation
it will be meaningless to talk of distinguishability, as one is not
experimentally distinguishing between the paths. For example, in a
two-path neutron interference experiment, if the spin state of the neutron
is different in each path, the triality relations (\ref{pceq}) and (\ref{pce})
will tightly quantify the relation between the predictability, visibility
(or coherence) and the measure of entanglement between the neutron paths
and the neutron spin. In this respect, entanglement can be considered
an integral part of wave-particle duality.

Several years back an interesting question was addressed in the context of
wave-particle duality. If the quanton is equipped with an internal degree
of freedom,  e.g., spin, and the interaction with the environment leads
to some path information being deposited in the environment, can one infer
the amount of this information from the interference visibility \cite{konrad}?
A duality relation was formulated for such a situation, using a much involved
``generalized visibility." \cite{konrad}. It is
easy to see that the triality relations (\ref{pceq}) and (\ref{pce})
can accomplish this task in a much simpler way. If one has knowledge of
the intensities of the two beams, and hence the predictability, 
measuring coherence from the interference will yield information about 
how much entanglement has been generated between the quanton and the
environment. These relations are also applicable to the already studied
problem of duality relation for a quanton in the presence of a quantum
memory \cite{buwu}.

These triality relations may also be used to \emph{measure} bipartite
entanglement in some experiments. If one wants to quantify the
entanglement between a quanton and an ancillary system, knowing the 
relative intensity of the two beams, and measuring coherence from the
interference will yield the measure of entanglement.

A recently introduced notion, in classical optics, was to connect the
wave-particle duality relation of the kind (\ref{GY}) to a generalized
version of \emph{polarization}. The result was the \emph{polarization
coherence theorem} \cite{eberly2}
\begin{equation}
P_F^2 = D^2 + V^2
\label{pct}
\end{equation}
where $P_F$ is the ``degree of polarization." It could represent polarization
due to the spin, or a ``mode polarization" related different spatial modes.
The quantity $D$ should be identified with predictability.
The triality relation (\ref{pce}) can be suggestively written as 
\begin{eqnarray}
 1 - \mathcal{E}^2 = \mathcal{P}^2 + \mathcal{C}^2,
\label{qpct}
\end{eqnarray}
where $\sqrt{1 - \mathcal{E}^2}$ plays the role of the ``generalized
polarization" $P_F$. In fact, for $n=2$ 
\begin{eqnarray}
1 - \mathcal{E}^2 = 1  - 4\Big[{\rho_r}_{11}{\rho_r}_{22} - 
|{\rho_r}_{12}|^2\Big] 
\end{eqnarray}
is closely similar to the expression of $P_F^2$ in Eq. (9) of
Ref. \cite{eberly2}.
This shows that the polarization coherence theorem has origins in the
entanglement between the quanton paths and some degrees of freedom
or ``modes." The expression $\sqrt{1 - \mathcal{E}^2}$, together with
Eq. (\ref{Esq}), may be explored further to understand higher dimensional
polarization coherence. 

We have looked at multipath interference of a quanton in the presence
of a path-detector. We find that the path-distinguishability is quantitatively
connected to path-predictability via entanglement. The well known \emph{duality}
relations between path-distinguishability and coherence can then
be viewed as \emph{triality} relations between path-predictability,
coherence and entanglement. We believe that these quantum mechanical results
lie at the root of a lot of recent interesting results in the realm
of classical optics. The works on ``completing" wave-particle duality
\cite{eberly3}, and ``turning off" duality \cite{eberly4}, are interestingly
connected to the relation between distinguishability and predictability.
The results obtained here also connect with the recently formulated
polarization coherence theorem \cite{eberly2}. One might wonder, how much
of quantum effects go into the phenomena that we have studied here. 
When the wave-particle duality is explored in the single photon regime,
then the entanglement between quanton paths and some other degrees of
freedom is genuine quantum entanglement.  At the other
end, in the classical optical regime, one may only talk of non-separability of
degrees of freedom. The fact the triality relation in the classical
optical regime (\ref{trialityC}) and that in the quantum regime 
(\ref{pce}), are very similar, indicates that non-separability may
be the only aspect of entanglement relevant here. 
We believe that this new picture in terms of the triality between
predictability, coherence, entanglement, opens the door to a wide
variety of interesting possibilities.

\medskip

\noindent\textbf{Acknowledgement.} 
The author thanks Joseph Eberly for bringing to his attention the role
of entanglement in the classical optical treatment of wave-particle duality.
He also thanks Xiaofeng Qian for suggesting possible role of entanglement
in classical optical multipath interference.

\medskip

\noindent\textbf{Disclosures.} The author declares no conflicts of interest.

\end{document}